\documentclass[conference]{IEEEtran}
\IEEEoverridecommandlockouts

\usepackage{cite}
\usepackage{amsmath,amssymb,amsfonts}
\usepackage{algorithmic}
\usepackage{graphicx}
\usepackage{textcomp}
\usepackage{xcolor}
\usepackage{subfigure}
\usepackage{comment}
\usepackage{bm}
\usepackage{caption}
\usepackage{amssymb}
\usepackage{calc}
\usepackage{url}
\def\BibTeX{{\rm B\kern-.05em{\sc i\kern-.025em b}\kern-.08em
    T\kern-.1667em\lower.7ex\hbox{E}\kern-.125emX}}
\begin{document}

\title{Investigation of perceptual music similarity \\focusing on each instrumental part\\
}

\author{\IEEEauthorblockN{Yuka Hashizume}
\IEEEauthorblockA{\textit{Graduate School of Informatics} \\
\textit{Nagoya University}\\
Nagoya, Japan \\
hashizume.yuuka@g.sp.m.is.nagoya-u.ac.jp}
\and
\IEEEauthorblockN{Tomoki Toda}
\IEEEauthorblockA{\textit{Information Technology Center} \\
\textit{Nagoya University}\\
Nagoya, Japan \\
tomoki@icts.nagoya-u.ac.jp}
}

\maketitle

\begin{abstract}
This paper presents an investigation of perceptual similarity between music tracks focusing on each individual instrumental part based on a large-scale listening test towards developing an instrumental-part-based music retrieval. In the listening test, 586 subjects evaluate the perceptual similarity of the audio tracks through an ABX test. We use the music tracks and their stems in the test set of the slakh2100 dataset. The perceptual similarity is evaluated based on four perspectives: timbre, rhythm, melody, and overall. We have analyzed the results of the listening test and have found that 1) perceptual music similarity varies depending on which instrumental part is focused on within each track; 2) rhythm and melody tend to have a larger impact on the perceptual music similarity than timbre except for the melody of drums; and 3) the previously proposed music similarity features tend to capture the perceptual similarity on timbre mainly.
\end{abstract}


\section{Introduction}
\label{sec:introduction}
Measuring music similarity is an important research topic due to its critical role in music retrieval and recommendation systems. Similarity that reflects human perceptual similarity is required for those systems to satisfy their users~\cite{Berenzwei2003,Ellis2002}.
However, it is difficult to label all of the huge music tracks available on the Internet manually. 
\par
In recent years, machine learning methods have been used to automatically calculate music features for unlabeled music tracks~\cite{Li2010,Hamel2010,mcfee2012,Lu2017,park2018,Lee2019,choi2019,Elbir2020,Clevelan2020,fatho2021}.  For example, features based on available attributes, such as genre~\cite{Elbir2020} and artist~\cite{park2018,Lee2019} have been developed. Also, we have proposed a method for learning the music similarity focusing on each instrumental part by metric learning~\cite{has2022, has2023}. This previous research aims to achieve flexible music retrieval and recommendation that allow users to choose the perspective to focus on. Our previously proposed method is a label-free training method to learn a similarity embedding space where the distance between different time segments within the same instrumental track tends to be small.
One limitation of this method is that a correspondence of the learned similarity to the perceptual similarity is unclear as the perceptual music similarity is not explicitly learned. 
\par
To investigate such correspondence, we also conducted subjective evaluation experiments through an ABX test with a small number of subjects in our previous study~\cite{has2022, has2023}. 
This evaluation method had some difficulties in handling sample sets with less agreement among subjects. The randomly selected reference and comparisons used in the ABX test had the potential to cause this issue because both comparisons were possibly similar/dissimilar to the reference to the same degree. One possible approach to address this issue could exclude answers with a low agreement rate among the subjects. However, it was actually hard to do that due to the following two reasons, 1) less agreement could be caused by the different evaluation criteria among the subjects, and 2) filtering out the low agreement answers resulted in a significantly small number of remaining answers. Consequently, these evaluations were insufficient to clarify the correspondence of the learned similarity to the perceptual similarity.
\par
This paper presents an investigation of perceptual similarity between music tracks focusing on each individual instrumental part based on a large-scale listening test. In the listening test, 586 subjects evaluate the perceptual similarity of the audio tracks for each instrument through the ABX test, and some of them also evaluate that for the mixed sound. We use the music tracks and their stems in the test set of the slakh2100 dataset~\cite{slakh}. The perceptual similarity is evaluated based on four perspectives: timbre, rhythm, melody, and overall. We analyze the results of the listening test to investigate the followings: 1) whether perceptual music similarity varies depending on which instrumental part is focused on within each track; 2) which perspective has a large impact on the perceptual music similarity, timbre, rhythm, or melody for each instrument; and 3) how much is the correspondence of the previously proposed music similarity features~\cite{has2023} to the perceptual music similarity. The answers collected in the ABX test are available as the dataset\footnote{\url{https://github.com/zume06/Inst-Sim-ABX-Dataset}}.

\section{Related works}
\label{sec:related_works}

\subsection{Research for perceptual music similarity}
Several studies have been conducted on perceptual music similarity~\cite{McAdams2001, Eerola2001, Berenzwei2003, Novello2006, Mullensiefen2007, Typke2005, Novello2011, Volk2012}. In the previous study~\cite{Ellis2002}, the authors asked subjects to evaluate the similarity between artists in popular music and collected 22,000 answers. They also conducted expert evaluations and analyzed the relationship between acoustic features and perceptual music similarity~\cite{Berenzwei2003}. Novello et al.~\cite{Novello2006} have analyzed the results of the similarity evaluation experiment for popular music in terms of consistency across subjects and so on. They have also researched the important factors underlying perceptual music similarity~\cite{Novello2011}. 
\par
These studies do not involve experiments where each individual instrumental part that composes a musical piece is listened to separately.

\subsection{Instrumental-part-based similarity learning}
\label{disent}
We have proposed a method to compute similarities focusing on each instrumental part with a single network that takes music tracks (i.e. mixed sounds) as input~\cite{has2023}. We have designed a similarity embedding space with disentangled subspaces for each instrument using Conditional Similarity Networks (CSNs)~\cite{CSN}.
\par
To disentangle the embedding space, a mask is applied to all dimensions except for the dimension corresponding to the notion to be considered in the triplet loss calculation. The network is given by function $f(\cdot)$, and $m_c$ is a mask that activates only the dimension corresponding to the condition $c$. The masked distance function between two samples $x_i$ and $x_j$ is given by
\begin {align}
\label{CSNtriplet}
d(x_i,x_j;m_c)=\parallel f(x_i)m_c-f(x_j)m_c\parallel_2.
\vspace{-24pt}
\end{align}
Letting $x_i^{(a)}$, $x_i^{(p)}$, and $x_i^{(n)}$ denote the $i$-th anchor, positive sample, and negative sample, respectively, $M$ denote the margin, the triplet loss can be written as follows.
\begin {align}
 &\mathcal{L}_T(x^{(a)},x^{(p)},x^{(n)},c)=\notag \\
 &\max\{0,d(x^{(a)},x^{(p)};m_c)-d(x^{(a)},x^{(n)};m_c)+M\}
\end{align}
\vspace{-24pt}
\\

 We have designed an embedding space whose subspaces are disentangled by the five instruments with condition $c$ where $c=0, 1, 2, 3, 4$ represent drums, bass, piano, guitar, and others, respectively. Let $D$ be the number of dimensions of subspace assigned for one instrument, and the subspace assigned to condition $c$ are as $f[cD:(c+1)D-1]$. The following formula defines the $m_c$ as a mask that leaves the subspace corresponding to each condition $c$ and sets the other dimensions to 0. 
\begin {align}
m_{ck} = \left\{
\begin{array}{ll}
1, & (cD\leqq k<(c+1)D)\\
0, & (\mbox{otherwise}).
\end{array}
\right.
\vspace{-8pt}
\end{align}

A triplet sampling method suitable for this method needs to satisfy a requirement: when learning a subspace corresponding to an instrument, an anchor and a positive sample are similar, and the anchor and a negative sample are dissimilar focusing on that instrument. To satisfy this requirement without using manual labels, for example in sampling for drums subspace, we have defined that music tracks that contain different time segments of the same drum sound are similar, and those that contain segments from different drum sounds are dissimilar. We have proposed the use of pseudo musical pieces and have shown that this method can improve the model’s performance.

\section{Perceptual similarity evaluation experiment}
\label{sec:experiment}
To investigate perceptual similarities focusing on each instrumental part, we performed ABX tests on the similarity of some sample sets of instrumental sounds and music tracks (i.e. mixed sounds).
\subsection{Experimental procedure for one sample set}
\label{sec:procedure}
Subjects were presented with three audio tracks of instrumental sounds, X, A, and B, and listened to all of them. They chose A+ if they perceived A to be more strongly similar to X than B, A- if slightly, B+ if they perceived B to be more strongly similar to X than A, and B- if slightly, based on the following four perspectives: timbre, rhythm, melody, and overall, respectively.
\par
In each answer, the subjects were allowed to select N/A from up to two perspectives except for \textit{overall}. The following instruction was provided with the subjects as cases where N/A can be selected; “A and B are similar/dissimilar to X of equal degree"; 
“The presented instrumental sound has no element corresponding to the perspective; e.g., drums have no melody." 

\begin{figure}
 \centerline{
 \includegraphics[width=1\columnwidth]{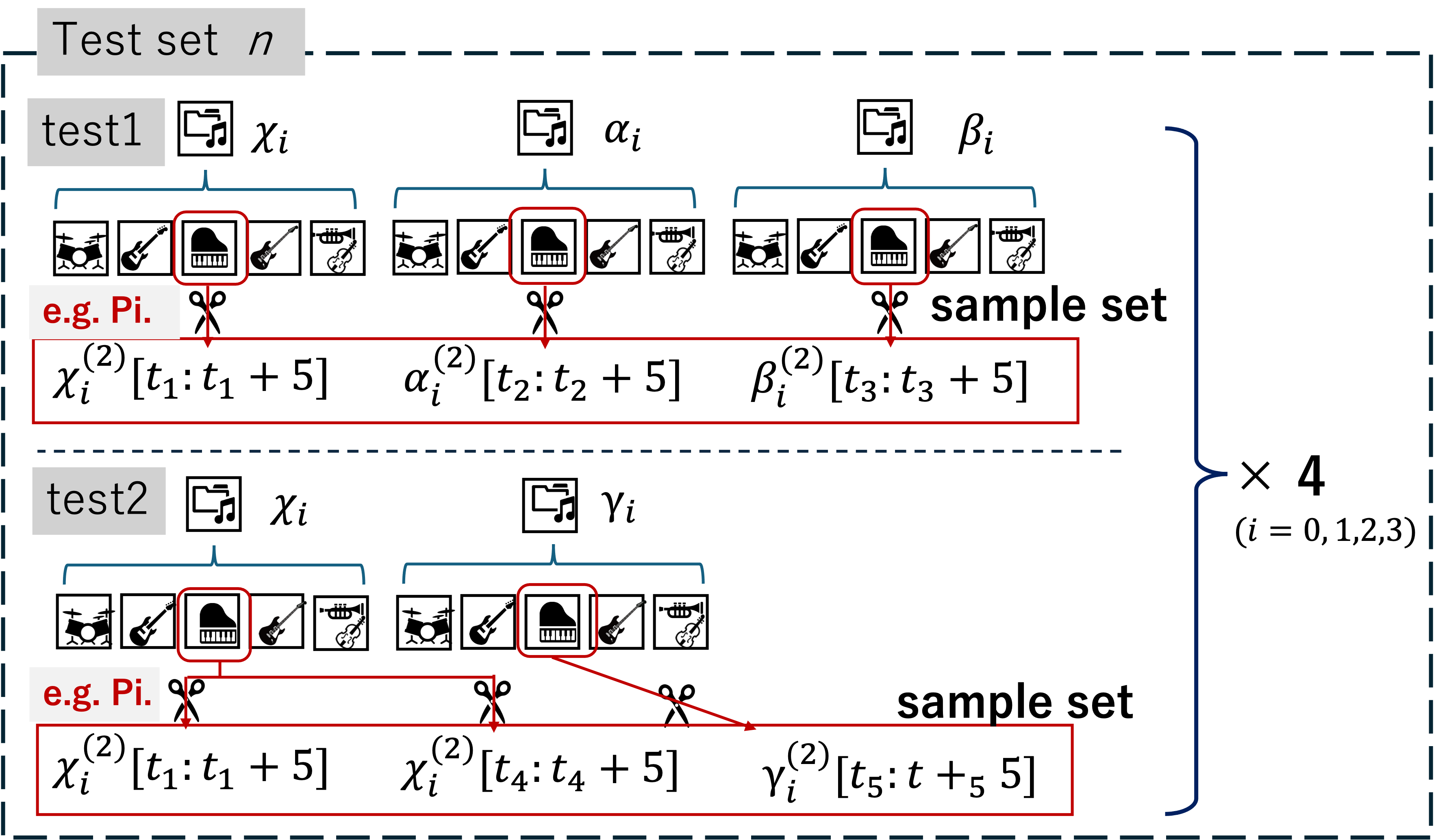}}
 \caption{How to create one test set. Examples of how to make one sample set for the piano sound are also shown with a red line for test 1 and test 2, respectively. Sample sets are created in the same way for other instrumental sounds (and the mixed sound for the additional experiment), and the procedure is repeated four times. }
 \label{fig:sample}
\end{figure}

\subsection{Sample selection}
\label{sec:selection}
 We made 2 types of sample sets, test 1 and test 2.
For test 1, we randomly selected three different music tracks $\{\chi_i, \alpha_i, \beta_i\}$ from the test set and randomly captured five-second segments from each instrumental sound contained in the three tracks, respectively. Then we obtained one sample set for each instrument, \{X, A, B\}=$\{\chi_i^{(j)}[t_1^{(j)}:t_1^{(j)}+5], \alpha_i^{(j)}[t_2^{(j)}:t_2^{(j)}+5], \beta_i^{(j)}[t_3^{(j)}:t_3^{(j)}+5]\}$. The subscript $j$ represents each instrument ($j=0,...,4$), with $0$ representing drums, $1$ bass, $2$ piano, $3$ guitar, and $4$ others. This selection was repeated four times $(i=0,...,3)$, and 20 sample sets were created.
\par
For test 2, we used the same X as in test 1, and one of the other two samples was taken from a different time of the same musical piece as X. Namely, we randomly selected a music track $\gamma_i$ from the test set excluding $\chi_i$, replaced $\alpha_i$, $\beta_i$ by $\chi_i$ and $\gamma_i$ (in no particular order), and the process was repeated in the same way as test 1. The 20 sample sets $\{\chi_i^{(j)}[t_1^{(j)}:t_1^{(j)}+5], \chi_i^{(j)}[t_4^{(j)}:t _4^{(j)}+5], \gamma_i^{(j)}[t_5^{(j)}:t_5^{(j)}+5]\}, (t_1 \neq t_4, i=0,...,3, j=0,...,4)$ were created, and it was randomly determined whether these set were \{X, A, B\} or \{X, B, A\}. Thus, a total of 40 sample sets were created, and these were used as one test set. An overview of the sample selection is shown in Fig.~\ref{fig:sample}. 
\par
In an additional experiment, 5-second segments were also extracted from each of the original music tracks (mixed sound) $\chi_i$, $\alpha_i$, $\beta_i$, and $\gamma_i$, and similar sample sets were created, i.e., 8 additional sample sets in total. Then, they were added to the test set, making a total of 48 sample sets as one test set. This procedure was repeated with a random selection of sample sets, creating a total of 60 test sets.

\subsection{Experimental settings}
\label{sec:setting}
The presented audio tracks were the 136 music tracks and the instrumental sounds that compose them. They were the tracks remaining from the 151 tracks in the test set of the redux subset of the slakh2100 dataset~\cite{slakh}, excluding tracks that do not contain enough instrumental sounds. The music tracks (mixed sounds) were used only in the additional experiment. 
\par
The ABX tests, consisting of 40 or 48 sample tests created as explained in Section~\ref{sec:selection} and 2 dummy ABX tests, were shuffled and presented to the subjects individually. Subjects were not informed whether the ABX test presented were test 1, test 2, or dummy. The dummy ABX tests were two types: A is the same audio as X, and B is the same audio as X.
\par
This research is aimed at establishing a system for use by general users rather than experts, experiments were conducted by recruiting participants through CrowdWorks ~\cite{cw}.

\subsection{Answer aggregation}
\label{sec:aggr}
\begin{figure}
\begin{tabular}{cc}
  \begin{minipage}[t]{0.9\hsize}
 \centering
 \includegraphics[width=0.9\columnwidth]{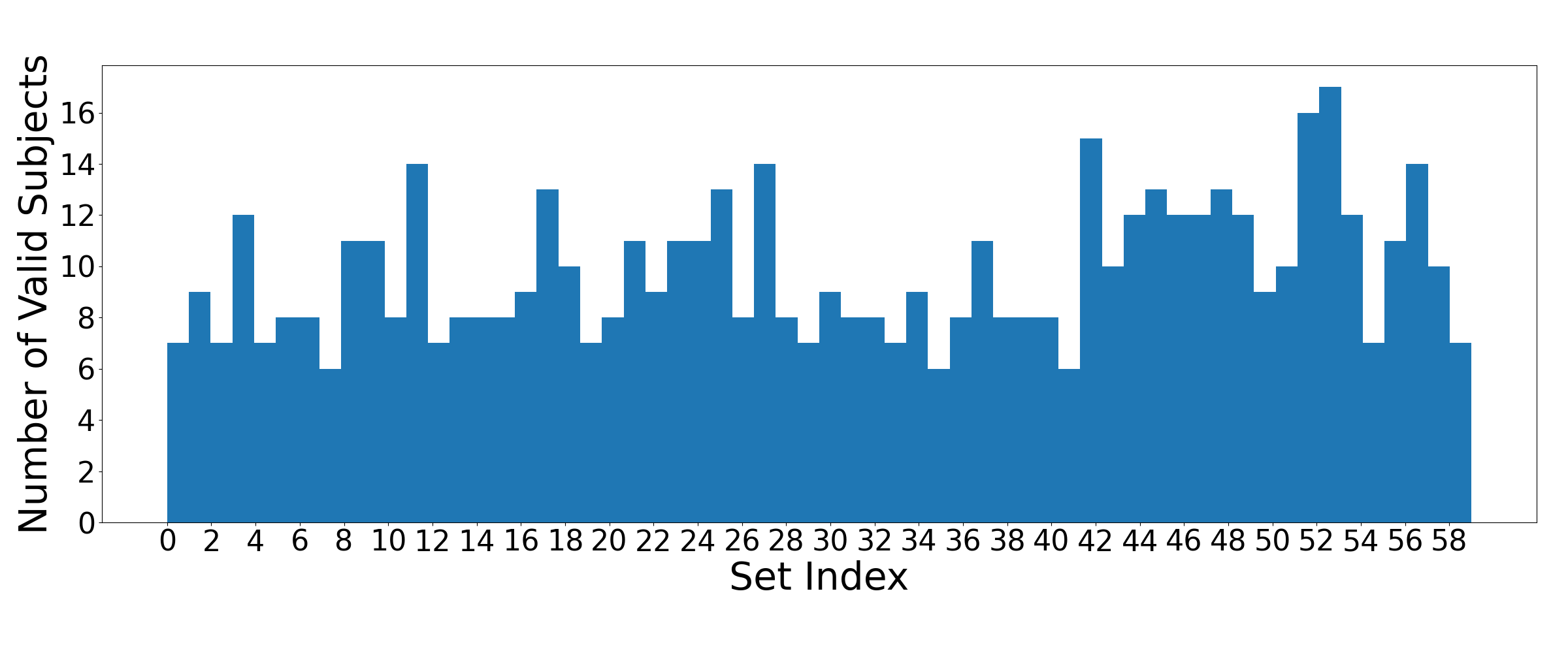}
 \vspace{-16pt}
 \caption*{(a) only instrumental sounds}
 \vspace{-8pt}
  \end{minipage}&
  \\
  \begin{minipage}[t]{0.9\hsize}
 \centering
 \includegraphics[width=0.9\columnwidth]{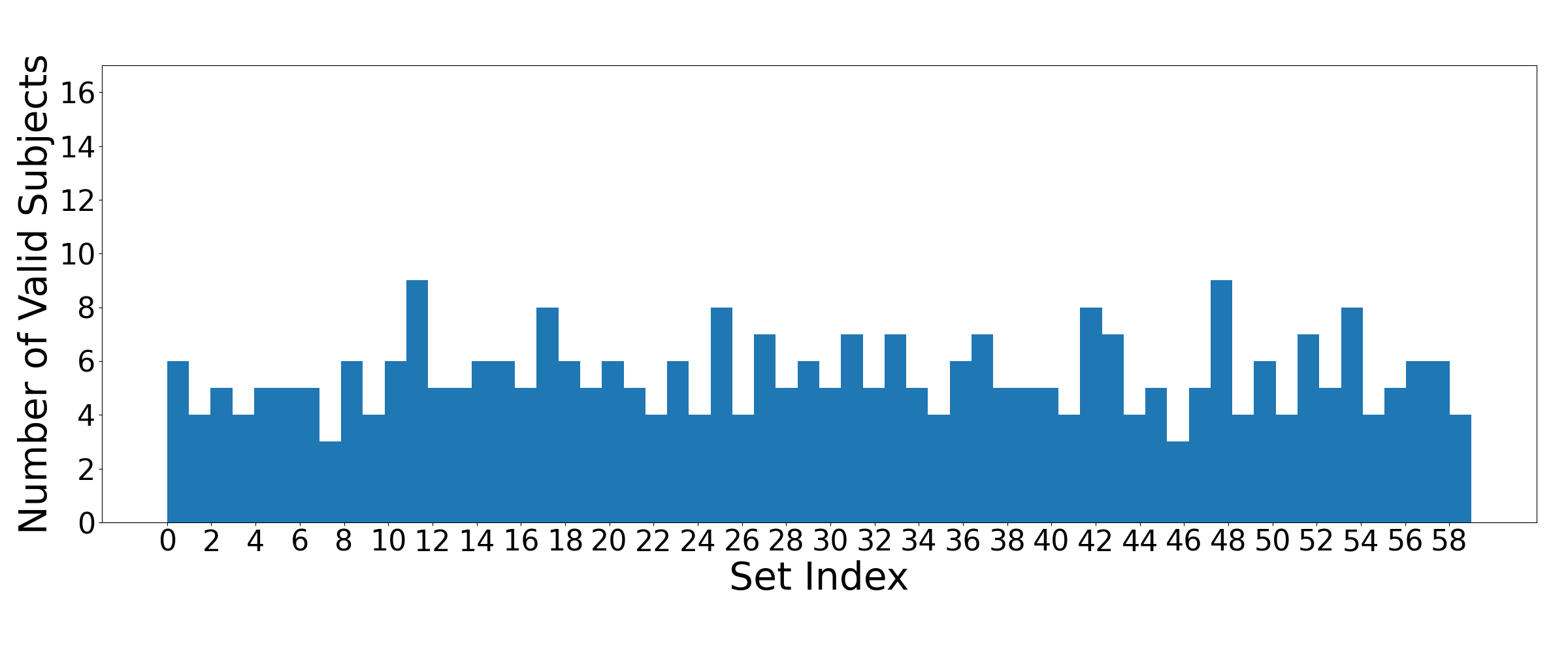}
 \vspace{-16pt}
 \caption*{(b) with mixed sounds}
 \end{minipage}
 \end{tabular}
 \caption{Histgrams of the number of subjects in each test set. The horizontal axis is the set index, and the vertical axis is the number of subjects who answered the corresponding set.}
 \label{fig:hist}
\end{figure}

After excluding all answers of subjects who did not select the same sample as X in the dummy tests, answers of other subjects that took less than 15 seconds including listening time, answers with blanks in the technical problems, and duplicate answers from the same subject in each test set were also excluded, then we obtained 26898 valid answers from a total of 586 subjects (281 unique subjects).
For all 60 sets, valid answers from at least 6 different subjects were obtained. Among them, the total number of valid subjects who also evaluated the music tracks (mixed sounds) in additional experiments was 328 (167 unique subjects). Fig.~\ref{fig:hist} shows a histogram of the number of subjects providing valid answers to each test set.

\section{Analysis}
In this paper, we define “A+" and “A-" are the same answer, and “B+" and “B-" are the same answer. In other words, the answer is treated as A, B, or N/A. 

\subsection{Differences in perceptual similarity between instruments}
For each subject, answers are obtained for each instrument for the four music sets $\{\chi_i, \alpha_i, \beta_i\} (i=0,1,2,3)$. For example, suppose that for one subject, the answer [$\alpha_0$, $\alpha_1$, $\beta_2$, $\beta_3$] is obtained for drums and [$\alpha_0$, $\alpha_1$, $\alpha_2$, $\beta_3$] for bass (subscripts for instruments are omitted). In this case, it means that in the music set $\{\chi_2, \alpha_2, \beta_2\}$, the subject perceives that the drum sound of $\beta_2$ is more similar to the drum sound of $\chi_2$, while the bass sound of $\alpha_2$ is more similar to the bass sound of $\chi_2$. Here, the matching rate between the drums and bass answers of this subject is 0.75. In this way, the matching rate between the answers of each instrument pair is calculated for each subject and averaged across all subjects.
\par
\begin{figure}
\begin{tabular}{cc}
  \begin{minipage}[t]{0.45\hsize}
 \centering
 \includegraphics[width=0.9\columnwidth]{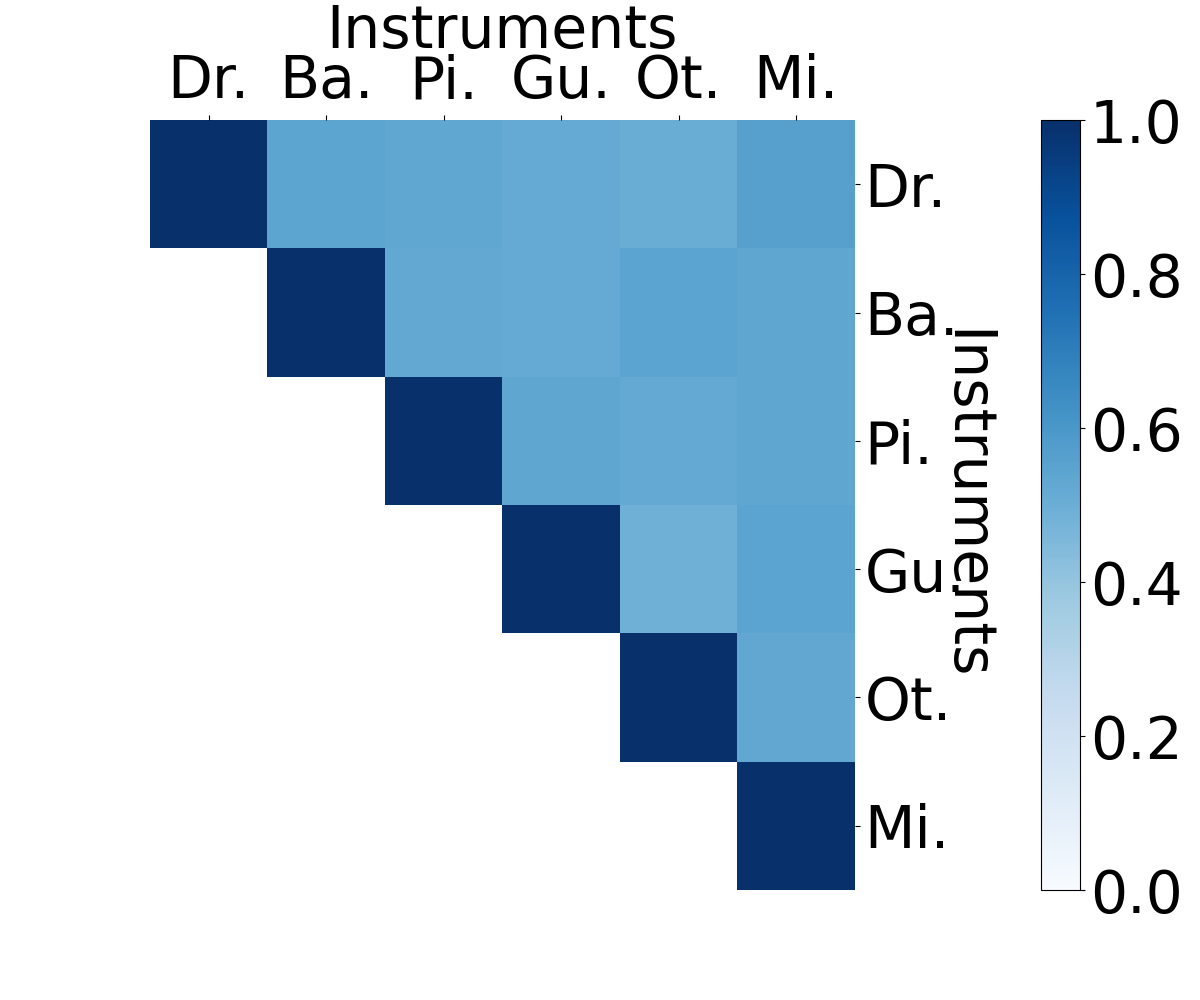}
 \caption*{(a) test 1}
  \end{minipage}&
  \hspace{1pt}
  \begin{minipage}[t]{0.45\hsize}
 \centering
 \includegraphics[width=0.9\columnwidth]{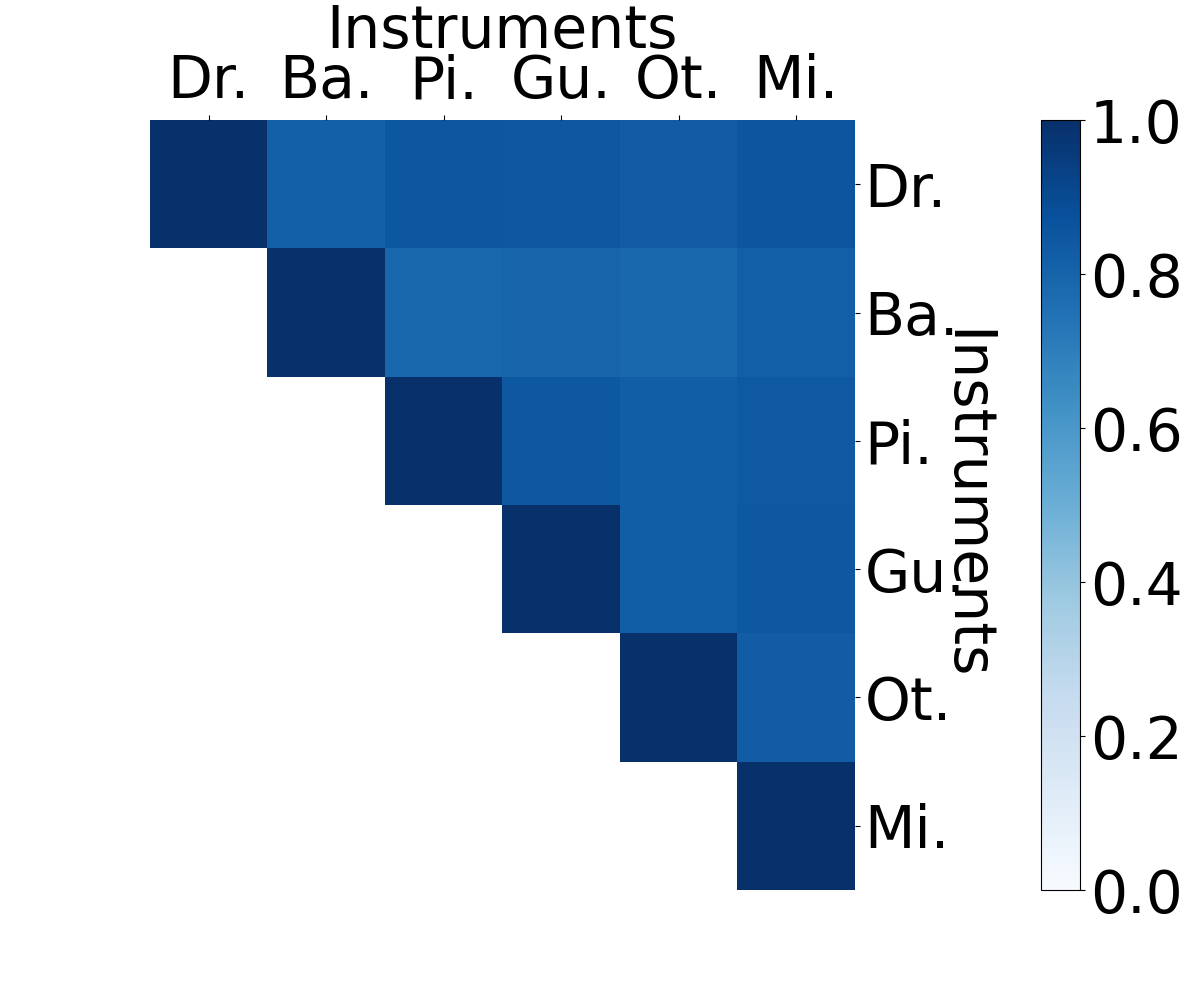}
 \caption*{(b) test 2}
 \end{minipage}
 \end{tabular}
 \caption{Heatmaps of the averages of matching rates of answers between instruments for a subject. “Dr.", “Ba.", “Pi.", “Gu.", “Ot." and “Mi." represent drums, bass, piano, guitar, others, and mixed sound respectively.}
 \label{fig:heatbwinst}
\end{figure}

The results are shown in Fig.~\ref{fig:heatbwinst}. In test 2, the matching rates between instruments is high because many subjects chose different segments of the same musical piece as X regardless of the type of instrument. In test 1, the values are comparable among all instruments, i.e., all of the off-diagonal values are around 0.5, indicating that the criteria for music similarity change depending on the instrumental part focused on. It is also suggested that no specific instrument has a consistent and significant influence on the similarity perception of musical pieces (mixed sounds). These results suggest that the function for searching and recommending similar music focusing on each instrumental part is effective.

\subsection{Impact of each perspective}
\label{sec:perspective}
As explained in Section~\ref{sec:procedure}, subjects evaluated music similarity based on four perspectives; timbre, rhythm, melody, and overall for one sample set. The subjects must choose either A or B in the \textit{overall}, even when the evaluation is divided among the three perspectives. By examining which perspective the subjects followed in selecting the \textit{overall}, we investigated which perspective is important for humans to listen to for each instrumental sound. 
\par
Matching rates of three perspectives with \textit{overall} are shown in Table~\ref{tbl:pers}. In drums, many of the answers for melody are N/A, and the matching rate with \textit{overall} for melody is low. For all instruments, the matching rate of timbre is as high as 70\% or more, but that of rhythm and melody are even higher than that of timbre except for drums. This means that rhythm and melody tend to have a larger impact on the perceptual music similarity than timbre in each instrumental sound, except for melody in drums.

\begin{table}[t]
\caption{Matching rate with \textit{overall} for each perspective (\%)}
\vspace{-8pt}
\label{tbl:pers}
\begin{center}
\scalebox{0.9}{
\begin{tabular}{c|cccccc}\hline
&drums & bass & piano & guitar & others & mix\\
\hline
Timbre&       74.7&         74.5&        75.6&          75.1&        75.3 &         74.3 \\
Rythm&\textbf{84.4}&\textbf{80.1}&       78.4&\textbf{80.6}&         79.8 &\textbf{82.0}\\
Melody&       56.7&79.0&\textbf{80.7}&\textbf{81.0}&\textbf{83.1}&\textbf{82.8}\\
\hline

\end{tabular}
}
\end{center}
\end{table}

\begin{table}[t]
\centering
\caption{Matching rate between the model's output and subjects' answers using results with 80\% agreement among subjects focusing on overall and timbre, respectively.}
\vspace{-8pt}
\begin{center}
\subtable[Evaluation with answers on \textbf{overall}]{
\scalebox{0.9}{
\begin{tabular}{c|ccccc}
      \hline
	&drums&	bass&	piano&	guitar&	others\\
 \hline
test 1	&52.7±4.9	&67.5±4.8	&58.5±5.1	&60.1±4.9	&58.3±4.8\\
test 2	&94.8±1.6	&88.6±2.4	&87.1±2.3	&92.0±1.9	&91.6±2.1\\
\hline
\end{tabular}}
}
\subtable[Evaluation with answers on \textbf{timbre}]{
\scalebox{0.9}{
    \begin{tabular}{c|ccccc}
      \hline

	&drums	&bass	&piano	&guitar	&others\\
 \hline
test 1	&\textbf{65.3±7.2}	&67.0±7.2	&\textbf{74.3±8.1}	&69.5±6.8	&59.5±7.7\\
test 2	&94.8±1.7	&86.5±2.6	&87.2±2.3	&92.0±2.0	&93.8±1.9\\
\hline
    \end{tabular}}
}
\end{center}
\vspace{-8pt}
\label{tab:eval_prev}
\end{table}

\begin{table}[t]
\centering
\caption{Number of answers using results with 80\% agreement among subjects focusing on overall and timbre, respectively.}
\vspace{-8pt}
\begin{center}
\subtable[Evaluation with answers on \textbf{overall}]{
\scalebox{0.9}{
\begin{tabular}{c|ccccc}
      \hline
	&drums&	bass&	piano&	guitar&	others\\
 \hline
test 1	&414&400&383&419&434\\
test 2	&965&965&920&934&850\\
\hline
\end{tabular}}
}
\subtable[Evaluation with answers on \textbf{timbre}]{
\scalebox{0.9}{
    \begin{tabular}{c|ccccc}
      \hline

	&drums	&bass	&piano	&guitar	&others\\
 \hline
test 1	&193&188&140&203&173\\
test 2	&879&879&913&892&834\\
\hline
    \end{tabular}}
}
\end{center}
\vspace{-16pt}
\label{tab:eval_prev_num}
\end{table}

\subsection{Evaluation of the method of the previous study}
We evaluated the method of the previous study~\cite{has2023} using answers obtained in this experiment. 
We calculated whether A or B was closer to X using the model~\cite{has2023} for the same set as used in this listening test and then calculated the matching rate between the model's results and the subjective evaluation results. The calculation of distances by the model was performed as follows: The music tracks originally containing the instrumental tracks used in the listening test were input into the model, and then the distance was measured by applying a mask that leaves only the subspace corresponding to the target instrument. As mentioned in Section~\ref{sec:introduction}, it is not suitable to use the sample sets in which the agreement rates among subjects are low. Only answers with larger agreement than 80\% among subjects were used in the evaluation. Note that if N/A accounts for the largest percentage, that sample set was not used in the evaluation (other than \textit{overall}).
\par
From Section~\ref{sec:perspective}, we found that there are several sample sets where the answers on timbre do not match the answers on overall. Thus, we evaluated the model using each of these answers.
The results are shown in Table~\ref{tab:eval_prev}, and the number of answers for each are shown in Table~\ref{tab:eval_prev_num}.
We can see that there is more agreement among subjects in the answers in test 2 than in test 1 from Table~\ref{tab:eval_prev_num}. Also, there are fewer answers for \textit{timbre} compared to \textit{overall} because the subjects can not select N/A in \textit{overall}, but they can in \textit{timbre}, and here N/A (after taking agreement) is omitted from the evaluation. As shown in Table~\ref{tab:eval_prev}, although test 1 is less accurate than test 2, accuracy is improved in drums, piano, and guitar in the evaluation using answers focusing on timbre compared to using answers focusing on overall. This suggests that the model is trained to represent similarity mainly focusing on timbre. 
\par
In conjunction with the finding in Section~\ref{sec:perspective} that subjects placed more importance on rhythm or melody than on timbre, we consider that if we can design a model that captures the structure of the time direction so that melody and rhythm can be considered, it will be possible to obtain a music similarity that is also compatible with human perception on overall.

\section{Conclusion}
This paper presents an investigation of perceptual music similarity focusing on individual instrumental parts in music tracks based on a large-scale listening test. We have analyzed the results of the listening test and have found that 1) perceptual music similarity varies depending on which instrumental part is focused on within each track, which suggests the significance of searching and recommending similar music for each instrumental part; 2) rhythm and melody tend to have a larger impact on the perceptual music similarity than timbre in each instrumental sound except for melody in drums. In addition, by increasing the number of subjects and by providing multiple perspectives to focus on, a more detailed evaluation of the model has provided the following findings: the previously proposed music similarity features tend to capture the perceptual music similarity on timbre mainly.
\section*{Acknowledgment}
This work was partly supported by JST CREST Grant Number JPMJCR19A3 and Grant-in-Aid for JSPS Fellows JP24KJ1253, Japan.

\vfill\pagebreak

\bibliographystyle{IEEEbib}
\bibliography{refs}

\end{document}